\documentclass[article]{ajs}

\usepackage{amsmath}
\usepackage{amsfonts}
\usepackage{amssymb}

\author{Omer Faruk Gulban\\ Maastricht University}
\title{The relation between color spaces and compositional data analysis demonstrated with magnetic resonance image processing applications}

\Plainauthor{Omer Faruk Gulban, First Author} 
\Plaintitle{The relation between color spaces and compositional data analysis demonstrated with magnetic resonance image processing applications} 
\Shorttitle{Colors and compositional data analysis} 

\Abstract{This paper presents a novel application of compositional data analysis methods in the context of color image processing. A vector decomposition method is proposed to reveal compositional components of any vector with positive components followed by compositional data analysis to demonstrate the relation between color space concepts such as hue and saturation to their compositional counterparts. The proposed methods are applied to a magnetic resonance imaging dataset acquired from a living human brain and a digital color photograph to perform image fusion. Potential future applications in magnetic resonance imaging are mentioned and the benefits/disadvantages of the proposed methods are discussed in terms of color image processing.}

\Plainkeywords{compositional data analysis, color, magnetic resonance imaging, image fusion} 

\Pages{1--xx}

\Address{
  Omer Faruk Gulban\\
  Cognitive Neuroscience Department, Faculty of Psychology and Neuroscience\\
  Maastricht University\\
  6224 EV Maastricht, Netherlands\\
  E-mail: \email{faruk.gulban@maastrichtuniversity.nl}\\
  URL: \url{https://orcid.org/0000-0001-7761-3727}
}

\begin{document}

\section{Introduction}

Compositional data analysis can be applied to color images in the context of image processing and analysis. Color images are stored as triplets of non-negative integers representing the additive primary colors red, green and blue (referred as RGB channels). Once the color image is formed, the transformation from RGB to hue, saturation and intensity (HSI) coordinate system is an often used first step to improve the image visualization \citep{Smith1978, Ledley1990, Grasso1993} (for the mathematical expression of this transformation see Appendix~\ref{RGB2HSI}). HSI color space is commonly used in computer applications because of the commonly accepted property of its components relating to human color perception \citep{Joblove1978, Levkowitz1993, Pohl2016}. Hue relates to the dominant component of the primary colors (red, green, blue) in the mixture, saturation relates to the distance from an equal mixture (gray), and intensity relates to distance from total darkness (black to white). In this article, inspiration is drawn from RGB to HSI color space transformation to propose an analogous and more general method based on compositional data analysis. During this process, a simple vector decomposition is outlined to justify the use of compositional data analysis \citep{Aitchison1982, Pawlowsky-Glahn2015} and the simplex space is leveraged to manipulate vector fields to perform image enhancement. The proposed method is demonstrated using a magnetic resonance imaging (MRI) dataset. The results are also visualized using a digital color photograph to provide a general intuition. Due to the interdisciplinary nature of this work a glossary is provided in ~Appendix~\ref{glossary} to accompany readers from different fields.

\section{Materials and Methods}

\subsection{MRI data acquisition and preprocessing} \label{data_and_preproc}

Whole head T1 weighted (T1w), proton density weighted (PDw) and T2* weighted (T2*w) images at 0.7 mm isotropic volumetric cubic element (voxel) resolution were acquired in one male participant using a three dimensional magnetization prepared rapid acquisition gradient echo sequence with a 32-channel head coil (Nova Medical) on a 7 Tesla whole-body scanner (Siemens) in Maastricht Brain Imaging Center. These measurements reflect different intrinsic properties of the tissues, for instance T1w image shows the optimal contrast between white matter and gray matter, PDw image shows the density of the hydrogen atoms, T2*w image shows the iron content. For the detailed report of the data acquisition parameters see ~Appendix~\ref{mri_parameters}. For a review on ultra high field MRI (($\geq7$ Tesla)) see \cite{Ugurbil2014}.

As a standard preprocessing step, a masking operation was performed based on the PDw image using FSL-BET software (version 2.1; \cite{Smith2002, Smith2004}). The resulting volumetric mask was applied to all images in order to discard parts containing non-brain tissues (e.g. bones, muscles, skin, air). This masking step was necessary to reduce the overall processing time in the following analyses by reducing the total number of voxels from 26214400 ($320\times320\times256$) to $\sim4.5$ million (4543582 to be exact). Measurements (T1w, PDw and T2*w) stored at each one of the $\sim4.5$ million voxels are considered as components of three separate scalar fields constituting a vector field when combined.

\subsection{Barycentric decomposition} \label{methods_bary_decomp}

The image analysis illustrated in this paper starts from the following vector decomposition, where the real space of $n$ dimensions is indicated with $\mathbb{R}^n$ and the simplex space is indicated with $\mathbb{S}^n$ symbols:
\begin{align} \label{bary_decomp}
  & \vec{v} = [v_1, v_2, \ldots, v_D] \text{, where } v_1, v_2, \ldots, v_D \in \mathbb{R}^D_{>0},\nonumber\\
  & \vec{v} = \frac{1}{k} C(\vec{v}) s \text{, where } C(\vec{v}) \in \mathbb{S}^D \text{, and } s \in \mathbb{R}^1_{>0},
\end{align}
where $\mathbb{R}^D_{>0}$ indicates positive real numbers, $k$ is an arbitrary scalar and the letter $C$ stands for the closure operation used in compositional data analysis \citep{Aitchison2002, Pawlowsky-Glahn2015}:
\begin{equation} \label{closure}
  C(\vec{v}) = k \frac{\vec{v}}{s}.
\end{equation}
The letter $s$ is another scalar that is the sum of the vector components:
\begin{equation} \label{intensity}
  s = \sum_{i=1}^D v_i.
\end{equation}
Note that $k$ disappears from the expression in ~Eq.~\ref{bary_decomp} and \ref{closure} when selected as one. The decomposition (Eq.~\ref{bary_decomp}) of the vector $\vec{v}$ into its barycentric coordinates ($C(\vec{v})$) and a scalar ($s$) allows the application of the compositional data analysis to the barycentric coordinates. Historically, the closure operation was used by August Ferdinand Mobius \citep{fauvel1993} and the resulting vector was interpreted as relating to the barycenter (center of mass) of a simplex, hence giving the name to the vector decomposition (Eq~\ref{bary_decomp}) demonstrated here.

\subsection{Compositional image analysis} \label{methods_coda}

In this section, operations performed on the the barycentric coordinates are laid out. These operations are mostly adapted from \cite{Pawlowsky-Glahn2015} to fit the context of the analyzed vector field. Tensor notation is used for explicit formulations:

Let $A$ be a tensor with three dimensions, $A_{x, y, z}$, where the coordinates $x, y, z$ of an element represents the spatial location. As mentioned the in Section \ref{data_and_preproc}, there are $\sim4.5$ million tensors (i.e. voxels) of interest in total. 

Let the superscript of tensor $A$ indicate different MRI measurements acquired at every voxel; $A_{x,y,z}^{T1w}$, $A_{x,y,z}^{PDw}$, $A_{x,y,z}^{T2^*w}$. As the first step, all three tensors are vectorized (flattened):
\begin{equation} \label{coda_start}
  \mathsf{vec}(A^t) = [a_{1,1,1},\ \hdots,\ a_{x,1,1}, a_{1,2,1},\ \hdots,\ a_{1,y,1}, a_{1,2,2},\ \hdots,\ a_{1,2,z}]^T.
\end{equation}
$T$ stands for the transpose operator and $ t \in T1w, PDw, T2^*w$.

Concatenation denoted by $\Vert$ is used to matricize the vectorized tensors as the next step:
\begin{equation}
  V = \mathsf{vec}(A^{T1w})\ \Vert \ \mathsf{vec}(A^{PDw})\ \Vert \ \mathsf{vec}(A^{T2w}),
\end{equation}
where number of rows of the matrix $V$ is $x \times y \times z$ ($n = \sim4.5$ million), and the number of columns is $3$. These operations are done for convenient indexing in the following equations.

Let $V$ indicate the set of voxels where the vector $v_i$ is the composition of T1w, PDw and T2*w measurements: 
\begin{equation}
  V = [v_i,\ \hdots,\  v_{n}] \text{ where } i \in [1, 2,\ \hdots,\ n] \text{ and } v_i = [v_{i}^{T1w}, v_{i}^{PDw}, v_{i}^{T2^*w}].
\end{equation}
At this stage, $V \in \mathbb{R}^3$. Voxel-wise (i.e. row-wise) closure is applied to $V$ to acquire the barycentric coordinates of every composition:
\begin{equation}  \label{barycentric_component}
  X = C(V) = [C(v_i),\ \hdots,\  C(v_n)],\ X \in \mathbb{S}^3.
\end{equation}
The set of compositions $X$ was centered by finding the \textit{sample center} and \textit{perturbing} each composition with the inverse of the sample center:
\begin{equation} \label{centering}
  \hat{X} = X \oplus \mathbf{\mathsf{cen}(X)^{-1}},
\end{equation}
where $\oplus$ denotes the perturbation operator (analogous to addition in real space):
\begin{equation}
  x \oplus y = C[x_{T1w}\mathbf{y}_1, x_{PDw}\mathbf{y}_2, x_{T2^*w}\mathbf{y}_3].
\end{equation}
Multipliers of the components are indicated with $\mathbf{y}_1$, $\mathbf{y}_2$ and $\mathbf{y}_3$. The term $\mathsf{cen}(X)$ is a vector that stands for the component-wise (i.e. column-wise) geometric mean  across all voxels:
\begin{equation}
  \mathsf{cen}(X) = [\mathbf{g}_{T1w}, \mathbf{g}_{PDw}, \mathbf{g}_{T2^*w}] = \left( \prod_{i=1}^{n} x_{ij} \right) ^ {1/n}, j = [T1w, PDw, T2^*w].
\end{equation}
After centering, the data is standardized:
\begin{equation}  \label{standardize}
  \hat{\hat{X}} = \hat{X} \odot \mathsf{totvar}[X]^{-1/2},
\end{equation}
where $\odot$ symbol stands for the power operator (analogous to scaling in real space). The exponent $p$ is applied to every component:
\begin{equation}
  x\odot p = C[x_{T1w}^p, x_{PDw}^p, x_{T2*w}^p],
\end{equation}
and the total variance is computed by:
\begin{equation}
  \mathsf{totvar}[X] =  \frac{1}{n} \sum_{i=1}^{n} d_a^2(x_i,\ \mathsf{cen}(X)),
\end{equation}
where $d_a^2$ indicates squared Aitchison distance:
\begin{equation}
  d_a(x, y) = \sqrt{\frac{1}{2D} \sum_{j=1}^D \sum_{k=1}^D \left( \ln\frac{x_j}{x_k} - \ln\frac{y_j}{y_k} \right)^2}.
\end{equation}
In the current example $D=3$ because of operating on a set of three part compositions $X$ (see ~Equation~\ref{barycentric_component}).

At this stage some visual intuition could be gained with regards to the vector field (MR images) that is decomposed and processed via compositional data analysis methods. For illustration purposes, this is done by generating virtual image contrasts computed through metrics of simplex space (see Figure~\ref{Fig2}).

The norm in $S^3$ image in Figure~\ref{Fig2} is generated by computing Aitchison norm voxel-wise:
\begin{equation}
  \parallel X \parallel _{a} = [ \parallel x_i \parallel _a,\ \hdots,\ \parallel x_n \parallel _a].
\end{equation}
The subscript $a$ stands for the vector norm defined in simplex space (analogous to Euclidean norm in real space). Following \cite{Pawlowsky-Glahn2015}, Aitchison norm is defined as:
\begin{equation} \label{a_norm}
  \parallel x \parallel _a = \sqrt{\frac{1}{2D} \sum_{j=1}^D \sum_{k=1}^D \left( \ln\frac{x_j}{x_k} \right)^2},
\end{equation}
where $D = 3$ considering three different MRI measurements (T1w, PDw, T2*w).

It should be noted that this norm image is analogous to saturation dimension in the aforementioned RGB to HSI color space transformation (see Appendix \ref{RGB2HSI}).

The angular difference in $S^3$ image in Figure~\ref{Fig2} is the voxel-wise angular difference ($\angle$) between the set of compositional vectors ($X$) and a reference vector ($r$):
\begin{equation}
  X_\measuredangle = [\angle{x_i r},\ \hdots,\ \angle{x_n r}],
\end{equation}
where
\begin{equation} \label{angular_diff}
  \angle{x_ir} = \arccos \left( \frac{\langle x_i, r \rangle _a}{\parallel x_i \parallel _a \parallel r \parallel _a} \right).
\end{equation}
$\parallel x \parallel _a$ stands for Aitchison norm of the vector $x$ as defined in ~Equation~ \ref{a_norm} and $\langle x, r \rangle _a$ stands for inner product in simplex space:
\begin{equation}
 \langle x, r \rangle _a = \frac{1}{2D} \sum_{j=1}^D \sum_{k=1}^D \ln\frac{x_j}{x_k} \ln\frac{r_j}{r_k}.
\end{equation}
Here, $D = 3$ again and it should be noted that the choice of the reference vector $r$ is arbitrary. In this example the reference vector is selected as $r = [0.05, 0.9, 0.05]$. This angle difference image is similar to hue dimension in RGB to HSI color space transformation (see Appendix \ref{RGB2HSI}).

Recognizing the similarity of norm and angular difference in simplex space to saturation and hue dimensions of HSI color space at this stage leads to the proposal of a color balance algorithm which is used together with centering (Eq. \ref{centering}) and standardization (Eq. \ref{standardize}) to generate the color image in Figure~\ref{Fig3} lower row:
\begin{equation}  \label{truncate}
  \mathsf{truncate}(x) = 
  \begin{cases}
  x \odot \left( \dfrac{\parallel x \parallel _a}{\lambda} \right) &, \parallel x \parallel _a > \lambda \\
  x &, \parallel x \parallel _a \leq \lambda
  \end{cases} 
\end{equation}
This method is inspired from other simple color balance methods that makes use of the dynamic range of RGB channels of color images \citep{Limare2011}. Truncate function is useful when when there are a small amount of outliers compositions. In essence this functions pulls the compositions that are too far away from the center of the simplex space towards the center. In the current work, distance threshold was arbitrarily chosen $\lambda = 3$ based on visual inspection of the resulting color enhancement. It is conceivable that other operations based on percentiles can also be used to the make the decision less arbitrary (e.g. $99^{th}$ percentile of Aitchison norm distribution consisting of all vectors of $X$).

The ilr transformation \citep{Egozcue2003,Pawlowsky-Glahn2015} was performed voxel-wise to acquire real space coordinates (ilr coordinates) of the set compositions to generate Figure~\ref{Fig3} right column:
\begin{equation}
  X_{\mathsf{ilr}} = [\mathsf{ilr}(x_i),\ \hdots,\ \mathsf{ilr}(x_n)],
\end{equation}
where the ilr transformation is defined as:
\begin{equation} \label{ilr_transformation}
  \mathsf{ilr}(x_i) = \ln(x_i) \cdot \mathbf{H}.
\end{equation}
$\mathbf{H}$ indicates the Helmert sub-matrix, chosen by following \cite{Tsagris2011,Lancaster1965}. In this case $\mathbf{H}$ consists of 3 rows and 2 columns and selected as the following:
\begin{equation} \label{helmert_matrix}
\mathbf{H} =
  \begin{bmatrix}
    \frac{1}{\sqrt{2}}  &  \frac{1}{\sqrt{6}} \\
    -\frac{1}{\sqrt{2}} &  \frac{1}{\sqrt{6}} \\
    0                   & -\sqrt{\frac{2}{3}}
  \end{bmatrix}
.
\end{equation}

To explore the usefulness of the ilr coordinates and probe physically interpretable compositional characteristics, a real-time interactive visualization method is used with joint ilr-coordinates and image space representations of the data. This method is similar to using pre-defined modulation transfer functions for mapping the bins of a 2D histogram data representation to 3D image space used in volume rendering software \citep{Kniss2005}. Major brain tissues such as gray matter, white matter, cerebrospinal fluid arteries and sinuses are identified manually by the author and delineated in both ilr-coordinates and brain images using neuro-anatomical expertise.

All analysis steps demonstrated in this work are implemented in a free and open source Python package (available at https://github.com/ofgulban/compoda; \cite{compoda0pt3pt1}) using Numpy \citep{numpy2011}, Scipy \citep{scipy2001}, Matplotlib \citep{matplotlib2007}, Nibabel \citep{nibabel2017} and scikit-image \citep{scikit-image} as auxiliary scientific libraries. The joint exploration of tissues in ilr-coordinates and image space is performed by using a specialized software developed by the author \citep{segmentator1pt3pt0}.

\section{Results}

\begin{figure}[hbt]
  \begin{center}
    \includegraphics[width=\textwidth]{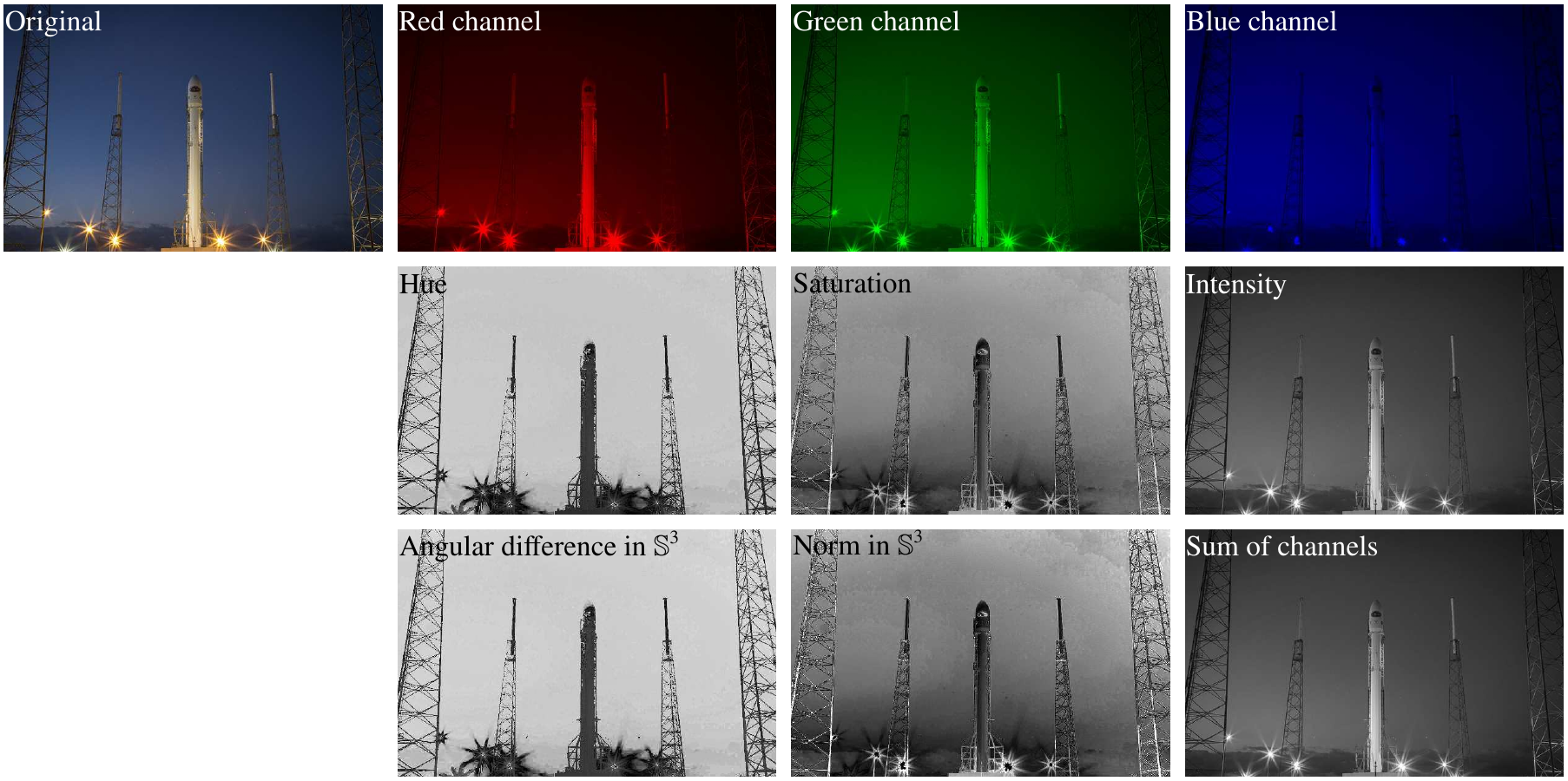}
    \caption{\label{Fig1}Visual demonstration of the similarity between RGB to HSI transformation and the corresponding analogous concepts laid out in ~Sections~\ref{methods_bary_decomp} and \ref{methods_coda}. In the first row the original image consisting of three channels and each of its channels can be seen. The second row shows hue, saturation and intensity components after RGB-HSI transformation [\ref{RGB2HSI}]. The third row shows angular difference in simplex space (analogous to hue), norm in simplex space (analogous to saturation) and derived by using compositional data methods (Equations~\ref{coda_start}-\ref{angular_diff} and the scalar component in real space (Equation~\ref{intensity}). Color maps of all scalar images range between the minimum and the maximum values of the corresponding image (darker shades are assigned to low values). The image is acquired from official SpaceX photo gallery (https://www.flickr.com/photos/spacex) licensed under CC0 1.0 public domain.}
  \end{center}
\end{figure}
Figure~\ref{Fig1} shows the similarity between HSI and the analogous metrics demonstrated in methods section. When comparing Fig.~\ref{Fig1} rows 2-3 column 1 to column 3, it can be seen that hue and angular difference in simplex space are both invariant to intensity gradient visible in the sky (brightness change from upper right corner to lower left). The same observation can also be made for the saturation and norm in simplex space images (compare Fig.~\ref{Fig1} rows 2-3 column 2 to column 3). These observations can be related to the scale invariance principle of compositional data analysis. By applying the barycentric decomposition (Eq.\ref{bary_decomp}) to color triplets in each pixel, scale information is separated, revealing compositional vectors invariant to the underlying multiplicative scalar field (i.e. brightness). These image contrasts based on compositional metrics are useful in object recognition tasks, see how the clouds are invisible in hue and angular difference images or the well-defined tip of the rocket in saturation and norm images.
\begin{figure}[hbt]
  \begin{center}
    \includegraphics[width=\textwidth]{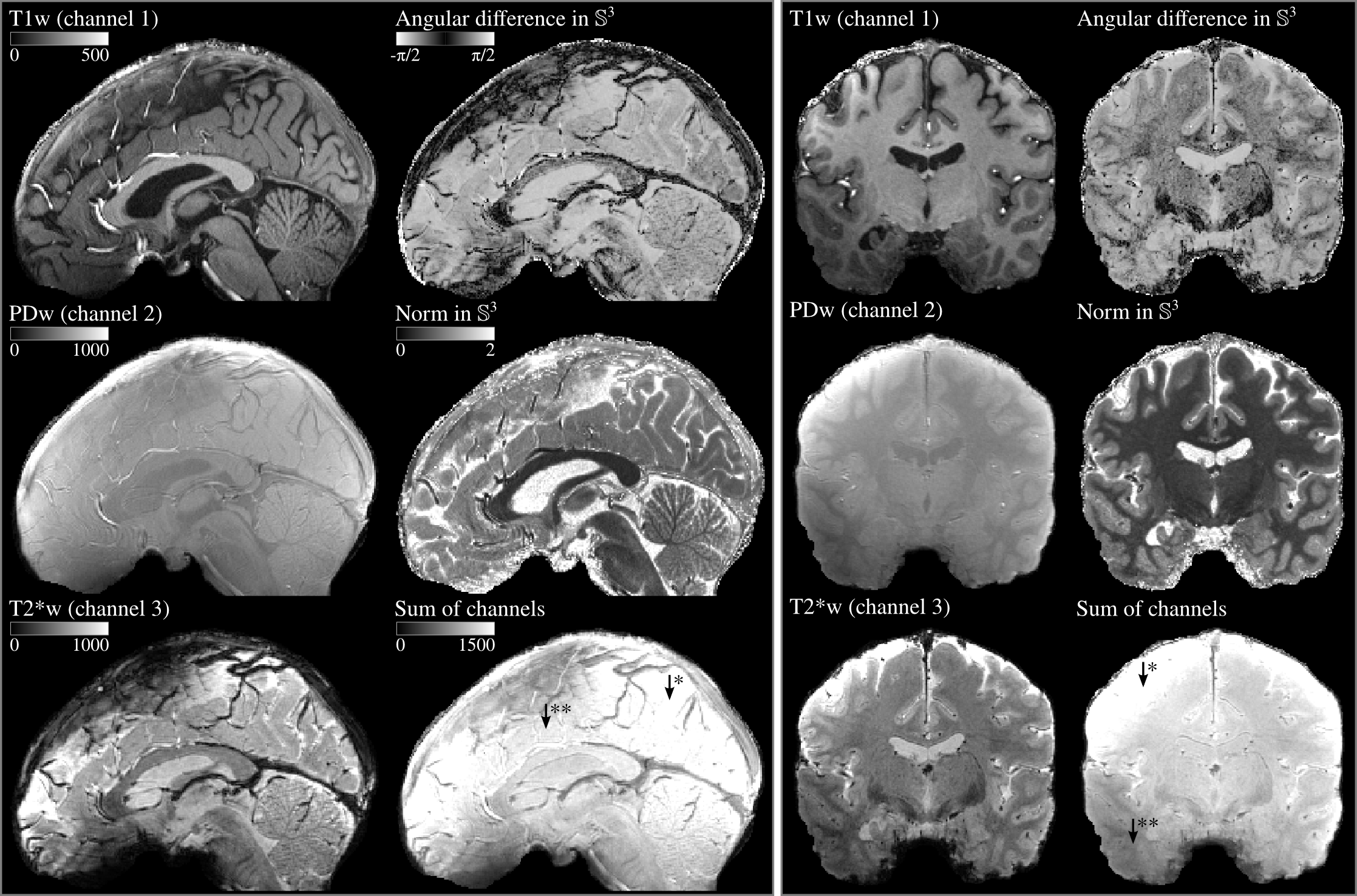}
    \caption{\label{Fig2}T1w, PDw and T2*w MR images rendered together with angular difference in simplex space (~Eq.~\ref{angular_diff}), norm in simplex space (~Eq.~\ref{a_norm}) and scalar channel sum component (~Eq.~\ref{intensity}). Left panel shows the sagittal slices (y-z plane) and the right panel shows a coronal slices (x-z plane) of the 3D images. Arrows with one asterisk (*) indicates the parts of the images that are too bright and arrows with two asterisks (**) indicates the parts that are too dim as a result of data acquisition imperfections. The legends on the upper left corner in left panel shows the colors bars which are also valid for the images in right panel.} 
  \end{center}
\end{figure}
Figure~\ref{Fig2} depicts two different slices of the 3D brain images visualized in gray scale depicting different image contrasts. It can be seen that the smooth, artefactual multiplicative scalar field (referred as bias field in MRI literature, see arrows with asterisks in Fig.~\ref{Fig2}) is separated from the compositional components (compare Fig.~\ref{Fig2} rows 1-2 column 2 with row 3 column 2 in both panels). This is similar to the separation of the brightness gradient in the sky in ~Fig.~\ref{Fig1} as mentioned in the previous paragraph. The compositional image contrasts can be considered as virtual contrasts which enhance the visual appearance by being invariant to certain artefactual features. It should be mentioned that other image contrasts can be generated by changing the reference vector in angular difference images (see Eq.~\ref{angular_diff}) to highlight specific tissues. Similarly, the compositions can also be perturbed (see Eq.\ref{centering}) differently to generate different contrasts in norm images. The generation of task-specific virtual contrasts could potentially be useful in medical imaging applications since the input channels and the virtual contrasts are physically interpretable.

\begin{figure}[hbt]
  \begin{center}
    \includegraphics{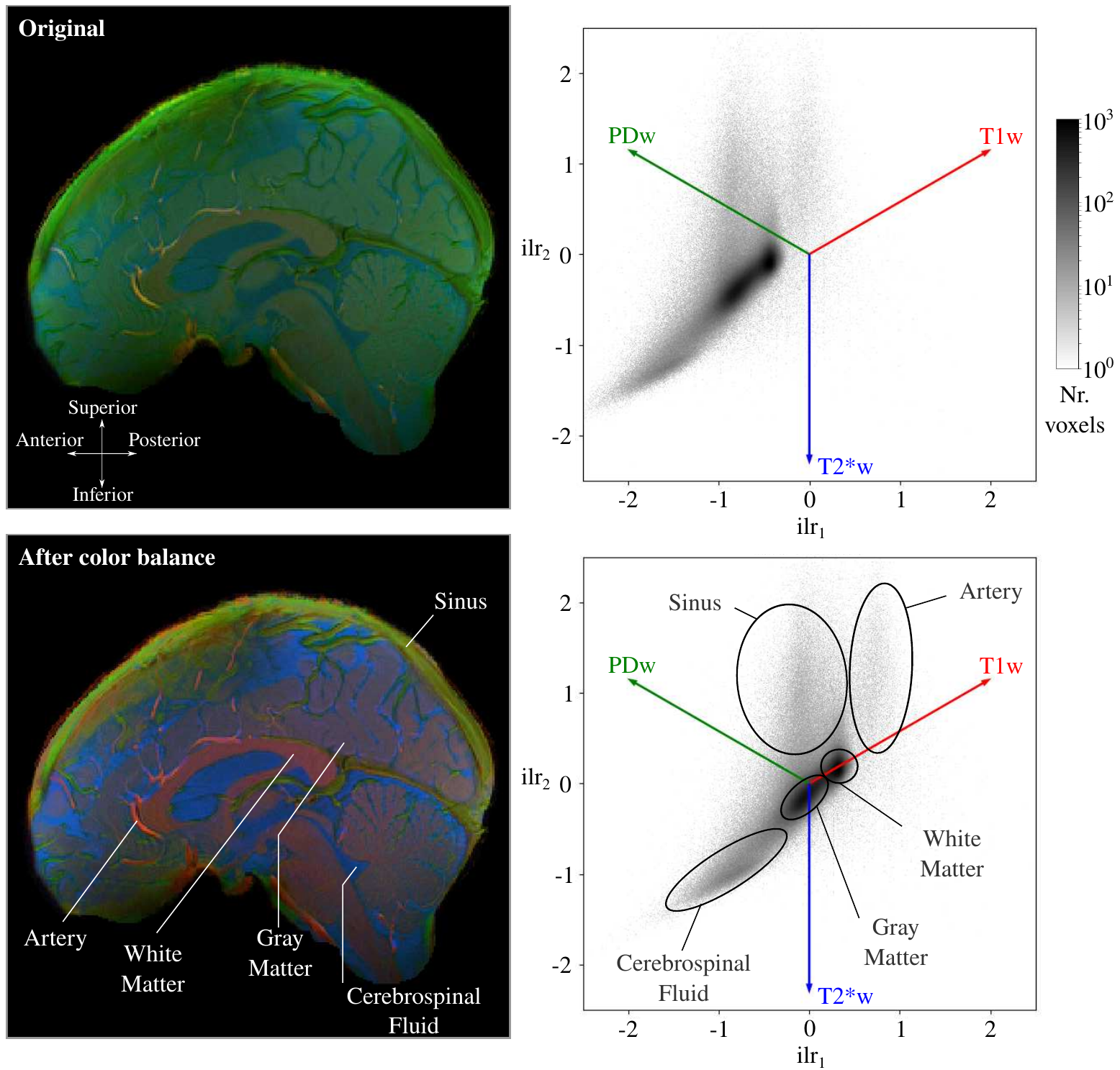}
    \caption{\label{Fig3}MRI measurements rendered as a color image. Red channel is assigned to T1w, green to PDw and blue to T2*w measurements. Left column shows the 2D histograms of the corresponding ilr coordinates (Eq.~\ref{ilr_transformation}). The projection of primary axes of RGB color cube (in $R^3$) to ilr coordinates are embedded to provide an intuitive reference for the characteristics of the compositions. The effect of centering and standardization inside the simplex (Eq. ~\ref{centering}, \ref{standardize}) and the compositional truncation (Eq. \ref{truncate}) is visible as color balance improvement in the rendered brain slice. The labels pointing to the tissues in brain image and circles in 2D ilr coordinate histograms shows the relation of the compositional characteristics with the coloration. The circles in 2D histograms are the edges of the 2D transfer functions used to manually probe the tissue-ilr coordinate relationship.} 
  \end{center}
\end{figure}

For the physical interpretation of the compositional vectors, a joint representation of color images and ilr coordinates of compositional vectors are presented in Figure~\ref{Fig3}. The difference between the rows in left column demonstrates the effect of color balance (see Equations~\ref{centering} and \ref{standardize}). It can be seen that the image contrast is enhanced to the level of easily recognizing major brain tissues by associating them to different colors. This is a more convenient visualization compared to inspecting T1w, PDw, T2*w image contrasts individually considering that the color image contains information from all three inputs at the same time. Right column in Fig.~\ref{Fig3} shows the 2D histograms of the ilr coordinates (see Eq.~\ref{ilr_transformation}) of the vector compositions. The effect of color balance becomes apparent when the compositions are interpreted with regards to the embedded RGB color cube primary axes (real space axes). For instance most of the clusters are on the left hand side of the center, close to the green arrow (PDw). This indicates that PDw measurements were initially dominating the compositions causing the green heavy color image. After the color balance, compositions spread more equally along the primary color axes, indicating that color contrast in the image will be richer (i.e. less dominated by a single component). The color balanced visualization is more useful for human observers in terms of intuitively recognizing the tissues. For instance the arteries have mostly reddish-white colors and the sinuses appear in green, which corresponds to the areas delineated for these tissues in ilr coordinates when the positions of clusters are considered relative to the embedded primary axes of RGB color cube. Although both of these are blood vessels, the difference between arteries and sinuses is meaningful because sinuses contain mostly deoxygenated hemoglobin, leading to a rapid decay of the MRI signal. In contrast, arteries contain oxygenated blood with slower MR signal decay. The difference in signal decay times of blood vessels effects T2*w and PDw measurements separating the compositional properties in relation of T1w images. Similarly the compositional change from white matter to gray matter to cerebrospinal fluid can be seen as an approximately straight line which corresponds to the change from red-heavy color of white matter to cyan of cerebrospinal fluid. It can be seen that these tissues are not dispersed towards PDw or T2*w axes, which can be intepreted as PDw and T2*w measurements not revealing different compositions in relation to T1w measurements when white matter, gray matter and cerebrospinal fluid is considered.

\section{Discussion}

In this work, compositional data analysis methods are used to reformulate RGB-HSI color space transformation. It is visually demonstrated that the compositional metrics such as angular difference and norm in simplex space relate to hue and saturation concepts in color space literature. This reformulation of hue and saturation would be advantageous for having a well-principled framework when operating on images with n-dimensions. For instance, a potential future application would be to analyze MRI datasets with more than three types of measurements by adding cerebral blood volume measurements \citep{Uludag2017} or multi-echo echo planar imaging \citep{Poser2009} to T1, T2 and PD weighted measurements at ultra high fields. Another application area would be processing of multi-spectral images for image fusion purposes \citep{Pohl2016}. In cases where more than three measurements are acquired for each element in an image, dimensionality reduction methods such as principal component analysis in simplex space \citep{Wang2015} would become relevant to maximize the visualized information content of color images. This is the disadvantage of being limited to three primary additive colors for color image rendering. However the virtual image contrasts (scalar images) could still be explored without dimensionality reduction. For instance the norm in simplex space can still be straightforwardly computed for n-dimensional compositions or the angular difference images can be explored can be explored by selecting different reference vectors to track the positions of n-dimensional compositions on an n-sphere. 

The disadvantages of compositional metrics presented here should also be mentioned. If the signal to noise ratio of the acquired images are low, compositional image contrasts would become less useful. For instance, the parts of images where only the measurement noise is recorded (e.g. thermal noise) the compositional metrics would carry no meaning. In such cases the angular difference and norm in simplex space would return enhanced noise patterns therefore the local brightness (i.e. intensity) should be taken into account before the compositional image analysis. The domain-specific exploration of the parameter spaces are also needed. For instance what is the extent of the usefulness of the compositional image processing when the noise properties are dissimilar between different types of measurements or when the measurements have different spatial scales (e.g. images with different resolutions). Further investigation is necessary to establish the relevance of the proposed application of compositional data analysis specifically to MRI data and generally to image processing and image fusion.


\section{Acknowledgements}

The data was acquired thanks to Federico De Martino, under the project supported by NWO VIDI grant 864-13-012. The author O.F.G. was also supported by the same grant. I thank Ingo Marquardt for language editing in the initial version of the manuscript and Alberto Cassese for the advice on mathematical notations in Section \ref{methods_bary_decomp}. In addition, I wish to express my appreciation for the comments and suggestions of the anonymous reviewers which I believe have helped to improve the manuscript.

\section{Appendix}

\subsection{Glossary} \label{glossary}

\textbf{Barycentric coordinates}: Center of mass-centric coordinates. Used in relation to the center of mass of an n-simplex.

\textbf{Color balance}: Used to indicate centering and standardizing compositional vectors in this work.

\textbf{HSI}: A three dimensional space with named axes relating to human color perception: hue, saturation, intensity.

\textbf{MRI}: Magnetic resonance imaging.

\textbf{PDw}: Proton density weighted MRI signal acquisition. Indicates density of the hydrogen atoms in different tissues.

\textbf{RGB}: A three dimensional space with named axes relating to color cube: red, green, blue

\textbf{Simplex}: Generalized geometrical notion of triangles. For example, 0-simplex is a point, 1-simplex is a line segment, 2-simplex is a triangle, 3-simplex is a tetrahedron etc.

\textbf{T1w}: T1 weighted MRI signal acquisition. Optimized to give the highest image contrast between white matter and gray matter brain tissues.

\textbf{T2*w}: T2* weighted MRI signal acquisition. Optimized to give the highest image contrast related to iron concentration between brain tissues.

\textbf{Voxel}: Volumetric cubic element, in other words a three dimensional pixel.

\subsection{RGB to HSI transformation} \label{RGB2HSI}

\begin{equation*}
  Intensity = I = \left(\frac{R}{\max(R,\ G,\ B)} + \frac{G}{\max(R,\ G,\ B)} + \frac{B}{\max(R,\ G,\ B)}\right) \div 3.
\end{equation*}

\begin{equation*}
Saturation = S =
  \begin{cases}
    0 &, I=0 \\
    1 - \dfrac{\min(R,\ G,\ B)}{I} &, I > 0.
  \end{cases}
\end{equation*}

\begin{equation*}
Hue = H =
  \begin{cases}
    0^\circ &, \Delta=0\\
    60^\circ \times \left( \dfrac{G' - B'}{\Delta}\mod6 \right) &, C_{max} = R'\\
    60^\circ \times \left( \dfrac{B' - R'}{\Delta} + 2 \right) &, C_{max} = G'\\
    60^\circ \times \left( \dfrac{R' - G'}{\Delta} + 4 \right) &, C_{max} = B'
  \end{cases}
\end{equation*}

\begin{equation*}
  \text{where } \Delta = \max(R,\ G,\ B) - \min(R,\ G,\ B)
\end{equation*}

\begin{equation*}
 \text{ and } R' = \frac{R}{\max(R,\ G,\ B)},\ G' = \frac{G}{\max(R,\ G,\ B)},\ B' = \frac{B}{\max(R,\ G,\ B)}.
\end{equation*}

\subsection{MRI data acquisition parameters and ethics statement} \label{mri_parameters}
Whole head images were acquired using a three dimensional magnetization prepared rapid acquisition gradient echo (MPRAGE) sequence. The data consisted of a T1w image (repetition time [TR] $= 3100\ ms$; time to inversion [TI] $= 1500\ ms$ [adiabatic non-selective inversion pulse]; time echo [TE] $= 2.42\ ms$; flip angle $= 5^\circ$; generalized auto-calibrating partially parallel acquisitions [GRAPPA] = 3 \citep{Griswold2002}; field of view [FOV] $= 224 \times 224\ mm^2$; matrix size $= 320 \times 320$; 256 slices; 0.7 mm isotropic voxels; pixel bandwidth $= 182$ Hz/pixel; first phase encode direction anterior to posterior; second phase encode direction left to right), a PDw image (0.7 mm isotropic) with the same 3D-MPRAGE sequence but without the inversion pulse (TR $= 1380\ ms$; TE $= 2.42\ ms$; flip angle $ = 5^\circ$; GRAPPA $= 3$; FOV $= 224 \times 224\ mm$; matrix size $= 320 \times 320$; 256 slices; 0.7 mm isotropic voxels; pixel bandwidth $= 182$ Hz/pixel; first phase encode direction anterior to posterior; second phase encode direction left to right), and a T2*w anatomical image using a modified MPRAGE sequence TR ($= 4910\ ms$; TE $= 16\ ms$; flip angle $= 5^\circ$; GRAPPA $= 3$; FOV $= 224 \times 224 mm$; matrix size $= 320 \times 320$; 256 slices; 0.7 mm isotropic voxels; pixel bandwidth $= 473$ Hz/pixel; first phase encode direction anterior to posterior; second phase encode direction left to right). Only magnitude images are stored after reconstruction.

The experimental procedures were approved by the ethics committee of the Faculty for Psychology and Neuroscience at Maastricht University, and were performed in accordance with the approved guidelines and the Declaration of Helsinki. Informed consent was obtained from the participant before conducting the data acquisition.

\bibliography{references}

\end{document}